\documentclass[journal]{IEEEtran}

\ifCLASSINFOpdf
\else
\fi

\usepackage[colorlinks, citecolor=blue]{hyperref}
\usepackage{balance}

\usepackage{setspace, amsmath, amssymb,subfig, url, lscape, algorithmic, multirow, pslatex, listings, verbatim, alltt, amsfonts, wrapfig, boxedminipage, color, cite,bookmark}
\usepackage[vlined,linesnumbered,ruled,boxed]{algorithm2e}
\usepackage[svgnames]{xcolor}
\usepackage{framed}
\definecolor{shadecolor}{named}{LightGray}

\usepackage{paralist}
\usepackage[dvips]{graphicx}
\usepackage{epsfig}
\let\labelindent\relax
\usepackage[inline]{enumitem}

\usepackage{xcolor}

\usepackage{listings}

\newcommand\YAMLcolonstyle{\color{red}\mdseries}
\newcommand\YAMLkeystyle{\color{black}\bfseries}
\newcommand\YAMLvaluestyle{\color{blue}\mdseries}

\makeatletter

\newcommand\language@yaml{yaml}

\expandafter\expandafter\expandafter\lstdefinelanguage
\expandafter{\language@yaml}
{
  keywords={true,false,null,y,n},
  keywordstyle=\color{darkgray}\bfseries,
  basicstyle=\ttfamily,                                 
  sensitive=false,
  comment=[l]{\#},
  morecomment=[s]{/*}{*/},
  commentstyle=\color{purple}\ttfamily,
  stringstyle=\YAMLvaluestyle\ttfamily,
  moredelim=[l][\color{orange}]{\&},
  moredelim=[l][\color{magenta}]{*},
  moredelim=**[il][\YAMLcolonstyle{:}\YAMLvaluestyle]{:},   
  morestring=[b]',
  morestring=[b]",
  literate =    {---}{{\ProcessThreeDashes}}3
                {>}{{\textcolor{red}\textgreater}}1     
                {|}{{\textcolor{red}\textbar}}1 
                {\ -\ }{{\mdseries\ -\ }}3,
}

\lst@AddToHook{EveryLine}{\ifx\lst@language\language@yaml\YAMLkeystyle\fi}
\makeatother

\newcommand{\qed}{\nobreak \ifvmode \relax \else
      \ifdim\lastskip<1.5em \hskip-\lastskip
     \hskip1.5em plus0em minus0.5em \fi \nobreak
      \vrule height0.75em width0.5em depth0.25em\fi}

\newcommand{\eg}{{e.g., }}

\newcommand{\comments}[1]{}
\newcommand\hl{\bgroup\markoverwith
  {\textcolor{yellow}{\rule[-.5ex]{2pt}{2.5ex}}}\ULon}

\usepackage{orcidlink}

\begin{document}

\title{Object as a Service: Simplifying Cloud-Native Development through Serverless Object Abstraction}
\author{
    Pawissanutt Lertpongrujikorn\,\orcidlink{0009-0003-4106-2347}, Mohsen Amini Salehi\,\orcidlink{0000-0002-7020-3810} 
     \\ High Performance Cloud Computing (\href{https://hpcclab.org}{HPCC}) Lab, University of North Texas
}


\captionsetup[subfloat]{captionskip=-0.5mm, farskip=-2mm}
\captionsetup[figure]{skip=0.6mm}

\maketitle

\IEEEpeerreviewmaketitle
\begin{abstract}
The function-as-a-service (FaaS) paradigm is envisioned as the next generation of cloud computing systems that mitigate the burden for cloud-native application developers by abstracting them from cloud resource management. However, it does not deal with the application data aspects. As such, developers have to intervene and undergo the burden of managing the application data, often via separate cloud storage services. To further streamline cloud-native application development, in this work, we propose a new paradigm, known as Object as a Service (OaaS) that encapsulates application data and functions into the cloud object abstraction. OaaS relieves developers from resource and data management burden while offering built-in optimization features. Inspired by OOP, OaaS incorporates access modifiers and inheritance into the serverless paradigm that: (a) prevents developers from compromising the system via accidentally accessing underlying data; and (b) enables software reuse in cloud-native application development. Furthermore, OaaS natively supports dataflow semantics. It enables developers to define function workflows while transparently handling data navigation, synchronization, and parallelism issues. To establish the OaaS paradigm, we develop a platform named \textit{Oparaca} that offers state abstraction for structured and unstructured data with consistency and fault-tolerant guarantees. We evaluated Oparaca under real-world settings against state-of-the-art platforms with respect to the imposed overhead, scalability, and ease of use. The results demonstrate that the object abstraction provided by OaaS can streamline flexible and scalable cloud-native application development with an insignificant overhead on the underlying serverless system.
\end{abstract}

\section{Introduction}\label{sec:intro}

\subsection{FaaS and Its Shortcomings}    
    
Function-as-a-Service (FaaS) paradigm is becoming widespread and envisioned as the next generation of cloud computing systems (a.k.a. Cloud 2.0) \cite{hassan2021survey} that mitigates the burden for programmers and cloud solution architects. Major public cloud providers offer FaaS services (\eg AWS Lambda, Google Cloud Function, Azure Function), and several open-source platforms for on-premise FaaS deployments are emerging (\eg OpenFaaS, Knative). FaaS offers the function abstraction that allows users to develop their business logic and invoke it via a predefined trigger. In the backend, the serverless platform hides the complexity of resource management and deploys the function seamlessly in a scalable manner. FaaS is proven to reduce development and operation costs via implementing scale-to-zero and charging the user in a pay-as-you-go manner. Thus, it aligns with modern software development paradigms, such as CI/CD and DevOps \cite{bangera2018devops}.

As the FaaS paradigm is primarily centered on stateless \emph{functions}, it naturally does not deal with \emph{data}. However, in practice, most use cases must maintain some form of state data. Moreover, FaaS limits direct communication between functions, making the application require additional remote storage for intermediate data. Thus, developers often have to intervene and undergo the burden of managing data using separate cloud services (\eg AWS S3~\cite{aws_s3}). For instance, in a video streaming application~\cite{msc}, developers must maintain video files, metadata, and access control in addition to developing functions.  
    
Apart from the lack of data management, current FaaS systems do not offer any built-in semantics to limit access to the functions' internal (private) mechanics. Providing unrestricted access to the developer team has known side effects, such as function invocation in an unintended context and data corruption via direct data manipulation. To overcome such side-effects, developers again need to intervene and undergo the burden of configuring external services (\eg AWS IAM \cite{aws_iam} and API gateway \cite{aws_api_gatway}) to enable access control.


Last but not least, current FaaS abstractions do not natively support function workflows. To form a workflow, the developer has to generate an event that triggers another function in each function. However, configuring and managing the chain of events for large workflows is cumbersome. Function orchestrator services (e.g., AWS Step Function\cite{aws_sf} and Azure Durable Function~\cite{azure_df}) can be employed to mitigate this burden. However, FaaS suffers from two inherent shortcomings: (a) limited direct communication between functions and (b) lack of built-in state management (see Figure~\ref{fig:intro}). These shortcomings force developers to intervene and employ other storage services to keep the intermediate data and manually navigate the data throughout the workflow~\cite{li2023dataflower}. In sum, although FaaS makes the resource management details (e.g., auto-scaling) transparent from the developer's perspective, it does not do so for data, access control, and workflow management.

\subsection{Proposed Paradigm}

\begin{figure}[tbp]
  \centering
  \subfloat[Function as a Service (FaaS)]{\includegraphics[width=0.36\textwidth]{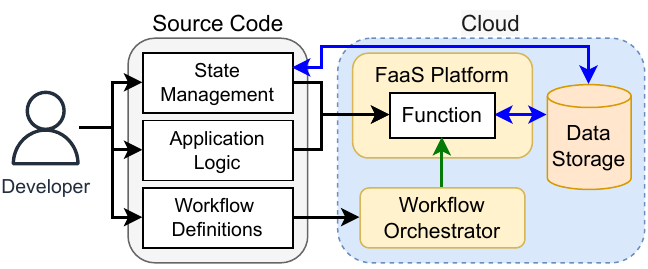}\label{fig:faas_cncpt}}
  \vspace{1mm}
  \hfill
  \medskip
  \subfloat[Object as a Service (OaaS)]{\includegraphics[width=0.42\textwidth]{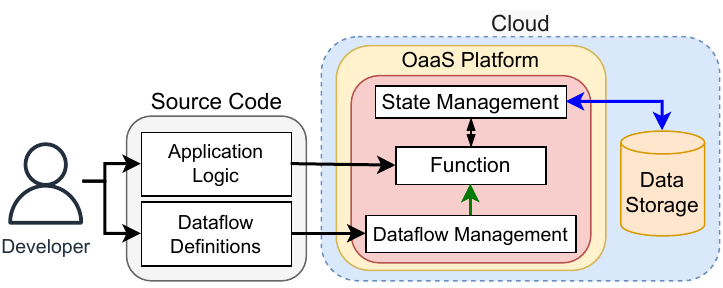}\label{fig:oaas_cncpt}}
  \caption{\small{A bird-eye view of FaaS vs. OaaS. 
  }}
  \label{fig:intro}
  \vspace{-6mm}
\end{figure}

To overcome these inherent problems of FaaS, we propose a new paradigm on top of the function abstraction that mitigates the burden of resource, data, and workflow management from the developer's perspective. \emph{We borrow the notion of ``object'' from object-oriented programming (OOP) and develop a new abstraction level within the serverless cloud, called \textbf{Object as a Service (OaaS)}} paradigm.

As shown in Figure \ref{fig:intro}, unlike FaaS, OaaS segregates state management from the developer's source code and incorporates it into the serverless platform, making it transparent from the developer's perspective. OaaS also incorporates workflow orchestration via its \textit{dataflow abstraction} with built-in data navigation across functions. Incorporating the application data and workflow into the object abstraction unlocks opportunities for built-in optimization features, such as data locality, data reliability, caching, as well as developer-guided hints to ensure Quality-of-Service targets~\cite{lertpongrujikorn2024streamlining}, enhance software reusability~\cite{denninnart2021harnessing}, and data access control. OaaS provides an interface for developers to declare the behavior and properties of objects in the form of \texttt{class} and \texttt{function}. For more complex use cases where the developer needs to orchestrate multiple objects and functions in the form of a workflow, OaaS offers the native (built-in) dataflow interface that can hide details of the data navigation and synchronization via declaring the flow of data in the form of an object's function.

An exemplar use case that can take advantage of OaaS is a cloud-based video streaming system (\eg \cite{msc,li2018performance}) that needs developers to rapidly implement new streaming services for the available video content. Examples of such services are blurring harmful (gore) content from videos, detecting faces on surveillance videos, and making the video processing pipeline for transcoding. Implementing these services using FaaS entails dealing with the state data (i.e., videos) and developing the business logic. In this scenario, the OaaS paradigm can mitigate the developer's burden by offering encapsulation. The videos are defined as persistent objects bound to a set of functions that can be invoked by the viewer's application and potentially change the object's (video's) state. 
For example, to blur gore segments in the video object \texttt{v1}, the developer invokes \texttt{v1.detect\_gore()} to obtain the output object \texttt{g1} that contains a list of time stamps of gore segments—next, blurring the video via invoking \texttt{v1.blurs(g1)}.

\subsection{Research and Contributions}



The \emph{\underline{first}} challenge is to establish OaaS by providing the “object” abstraction while ensuring the platform’s modularity and extensibility. The primary requirement is to offer an interface for developers to declare object behavior and state as classes, functions, and dataflow. With the addition of access modifiers, OaaS can encapsulate internal mechanics, allowing objects to reference other objects and form dataflow functions. This high-level abstraction enforces strong encapsulation, preventing invalid access and thereby unraveling complexity. To realize the OaaS paradigm, we develop \textbf{Oparaca} (\underline{\textbf{O}}bject \underline{\textbf{Para}}digm on Serverless \underline{\textbf{C}}loud \underline{\textbf{A}}bstraction), that is driven by flexibility and modularity. It must accommodate multiple use cases, remain extensible, and integrate seamlessly with various execution (FaaS) and storage modules. By leveraging pure functions, Oparaca avoids tight coupling with execution modules and offloads tasks without side effects. A data tiering scheme further abstracts storage, ensuring that changes in storage types do not necessitate modifications in the function code.

The \emph{\underline{second}} challenge is to reduce the overhead of data movement between functions and the platform's components. For that purpose, the data tiering scheme within the Oparaca platform diminishes the latency of accessing the object by employing a distributed in-memory hash table~\cite{yang2021large}. However, for unstructured data, also known as BLOB (e.g., multimedia content), that cannot fit in memory, Oparaca persisted them in a separate object storage. To further mitigate the overhead of accessing objects, Oparaca is equipped with a presigned URL with a redirection mechanism that reduces unnecessary data movements within the platform instead of relaying (i.e., transferring) the object state.



The \emph{\underline{third}} challenge is to ensure fault tolerance and data consistency. Concurrent requests to the same state can lead to data inconsistencies due to race conditions, and system failures may result in mismatched states across multiple data stores. Oparaca introduces a \textit{fail-safe state transition} mechanism that maintains data consistency and fault tolerance to tackle these issues. Additionally, Oparaca combines lightweight optimistic locking~\cite{leis2019optimistic} and \textit{localized locking} to prevent concurrent modifications to the same object. Another aspect of fault tolerance is the recovery mechanism from failure via retrying. However, such a remedy can potentially lead to another problem---repeating the execution more than once and falling into an undesirable state. To cope with this challenge, we develop a mechanism within Oparaca that establishes consistent state transitioning by guaranteeing ``exactly-once" execution for the function calls.


 In summary, the key contributions of this research are as follows:
\begin{enumerate} [leftmargin=*]
    \item Proposing A novel 
    \textbf{OaaS} paradigm that simplifies cloud-native development by unifying application data, business logic, and workflows into a single object abstraction. Uniquely, OaaS applies principles from OOP to the serverless model by introducing cloud class inheritance, which enhances software reusability across various cloud functions and runtimes.
    \item Developing \textbf{Oparaca}\footnote{The source code, packages, and example use cases are available here: \url{https://github.com/hpcclab/OaaS}}, as a prototype implementation of OaaS. Oparaca supports both structured and unstructured data while ensuring data consistency through a fail-safe state transition mechanism and a scalable, localized locking scheme to prevent race conditions without the overhead of cluster-wide coordination
    \item Devising a declarative dataflow abstraction that allows developers to define complex function workflows based on data dependencies rather than task dependencies. The Oparaca platform makes this high-level abstraction practical by using efficient techniques, such as presigned URLs with redirection and distributed in-memory caching, to minimize performance overhead and improve scalability compared to traditional FaaS state management.
    \item Analyzing Oparaca's performance from the scalability, overhead, and ease-of-use perspectives.
\end{enumerate}

\vspace{-1mm}
\section{Background and Prior Studies}
\label{sec:bck}

 

\begin{figure}[tbp]
  \centering
  \includegraphics[width=0.48\textwidth]{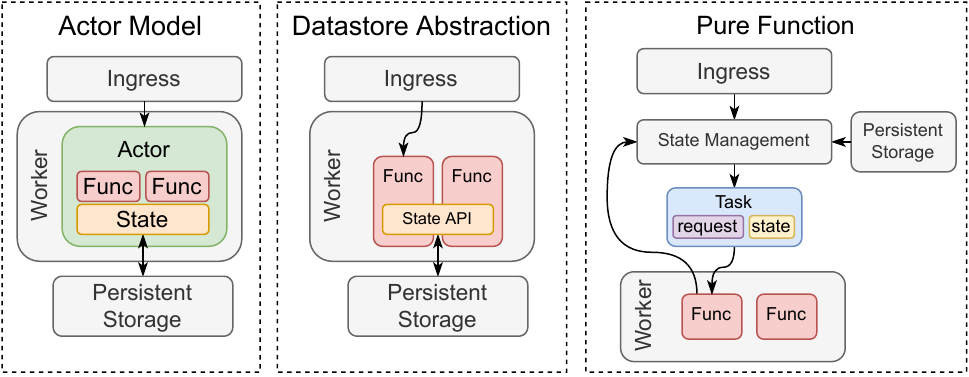}
  \caption{\small{The illustrated comparison of three different models of stateful serverless.}}
  \vspace{-2mm}
  \label{fig:stateful_serverless}
\end{figure}

The idea of stateful serverless is explored in several research works. These approaches primarily aim to address the stateless nature of the FaaS model, where the burden of managing application data, access control, and function workflows is often shifted to the developer through separate cloud services. As noted in Figure~\ref{fig:stateful_serverless}, these works can be categorized into actor model, datastore abstraction, and pure function approaches depending on how the platform manages data and allows functions to access it.

\noindent\textbf{Actor Model}. In the actor model, the platform fetches the state from persistent storage and places (i.e., caches) it inside a worker instance, then dispatches the request to where the state resides to achieve data locality. This approach, however, can make maintainability difficult for bulky unstructured data. The deployment granularity is an actor that contains multiple functions sharing the same state. The foundational platform in this space is Orleans~\cite{bernstein2014orleans}, which introduces ``virtual actors". Its influence is evident in modern platforms such as Kalix~\cite{kalix} and Azure Entity Functions~\cite{azure_enfunc}. While effective for certain use cases, this model can tightly couple state and compute within a single language and environment. In contrast, Oparaca manages the object abstraction at the platform level, allowing functions implemented in different languages or runtimes to operate on an object, thereby offering greater flexibility.

\noindent\textbf{Datastore Abstraction}. The datastore abstraction is a hybrid approach where the platform provides an API for the function to access data on demand. Like pure functions, it relaxes the need for state and function co-location, but can utilize caching to preserve data locality. Several systems utilize this pattern. Cloudburst~\cite{cloudburst} uses a shared distributed key-value database, while Boki~\cite{jia2021boki} provides API access to a distributed logging system. This log-based approach for consistency differs from Oparaca's mechanisms, which use a combination of fail-safe state transitions for unstructured data and localized locking to handle race conditions. Beldi~\cite{beldi}, on the other hand, provides the database and transaction API to the state. While providing direct transactional APIs offers strong guarantees, Oparaca abstracts these concerns from developers through its fail-safe versioning scheme and guarantees of exactly-once execution, aiming for a higher-level resilience model.
FAASM~\cite{faasm} optimizes the function-state interaction by using WebAssembly~\cite{wasm}. Although this enables multiple functions to share memory and achieve data locality, it requires compiling code to WebAssembly, which can limit library compatibility. Oparaca’s hybrid model avoids this limitation by combining the pure function and datastore abstraction approaches, offering broader language support.

\noindent\textbf{Pure Function}. In the pure function approach, the state is placed on another system and transferred to the worker instance upon invocation, appearing as part of the function's input arguments. This disaggregates state management and computation for system design simplicity, but can compromise data locality. Apache Flink Stateful Function (StateFun)~\cite{statefun} and Kalix~\cite{kalix} are solutions based on this approach. Oparaca combines the pure function approach for structured data with a lazy-fetch mechanism for unstructured data, practically employing both the pure function and datastore abstraction approaches.

\noindent\textbf{Process Abstractions}. While OaaS simplifies development by \textit{increasing} abstraction, a competing trend in recent work involves stepping back from abstraction to grant developers more control for performance. This trend of using OS-inspired abstractions is evident in various contexts. For serverless platforms, Process-as-a-Service (PaaS)~\cite{copik2024process} introduces a \textit{cloud process} to optimize execution, while for traditional always-on services, Nu~\cite{ruan2023nu} utilizes migratable \textit{logical processes}.

\noindent\textbf{Serverless Workflow and Dataflow} Beyond state management for individual functions, composing them into complex applications is another significant challenge. Commercial orchestrators, such as AWS Step Functions and Azure Durable Functions, can mitigate the burden of chaining events; however, developers are often still required to navigate data between steps manually. The research community has proposed more integrated solutions. For instance, DataFlower~\cite{li2023dataflower} introduces a dataflow paradigm specifically for serverless workflow orchestration to optimize data movement between functions. Netherite~\cite{azure_df} focuses on the efficient execution of these workflows through techniques like partitioned state management and collocated execution. While these systems significantly advance workflow performance and orchestration, OaaS approaches the problem from a different angle by integrating workflow directly into its core object abstraction.

Oparaca’s dataflow functions allow developers to declaratively define a workflow as a DAG based on data dependencies rather than task dependencies. The platform transparently manages parallelism, data navigation, and state consistency, abstracting these complexities from the developer.





\section{Object as a Service (OaaS) Paradigm}\label{sec:oaas}

\subsection{Conceptual Modeling of OaaS}
To realize OaaS, \textit{\underline{first}}, we need to establish the notion of \emph{cloud object} as an entity that possesses a \emph{state} (i.e., data) and is associated with one or more \emph{functions}. We empower objects to support both structured (\eg JSON records) and unstructured (\eg video) forms of state. Upon calling an object's function, OaaS creates a task that can safely take action on the state. 

\textit{\underline{Second}}, OaaS provides the \emph{class} semantic as a framework to develop objects. Inspired by OOP, the developer has to define a set of functions and states within the class. Then, an arbitrary number of objects---that is bound to the functions and states declared in that class---can be instantiated. To improve cloud software reusability and maintainability, we enable class \textit{inheritance} for cloud functions and states from other classes, plus the ability to \textit{override} any derived function. 

\textit{\underline{Third}}, OaaS offers built-in \textit{access control} to provide the ability to declare the ``scope of accessibility'' for a state or function. Importantly, when defining a set of classes, the developer can declare it within a single package that includes the access modifier to prevent unauthorized access from other packages. This is particularly useful when cloud application developers utilize imported third-party packages.

\textit{\underline{Fourth}}, OaaS enables higher-level abstractions by allowing cloud objects to be nested, where a high-level object references lower-level objects. Functions can use these references to fetch inputs or invoke dataflow functions (called \textit{macro functions}) that chain operations across lower-level objects. Unlike traditional FaaS workflows \cite{aws_sf}, macro functions determine execution flow based on dataflow rather than task (i.e., function call) dependency. Developers only define the data flow, while OaaS manages parallelism, data navigation, and state consistency transparently.

\subsection{Developing Classes in OaaS}

In OaaS, developers define one or more classes within a package using configuration languages like YAML or JSON. The package definition contains the class section and the function section. The functions section defines the configuration and deployment details of each function. The class section defines the object's structure, which includes the state and function it links to.

\begin{figure} [t]
  \centering
\includegraphics[width=0.49\textwidth]{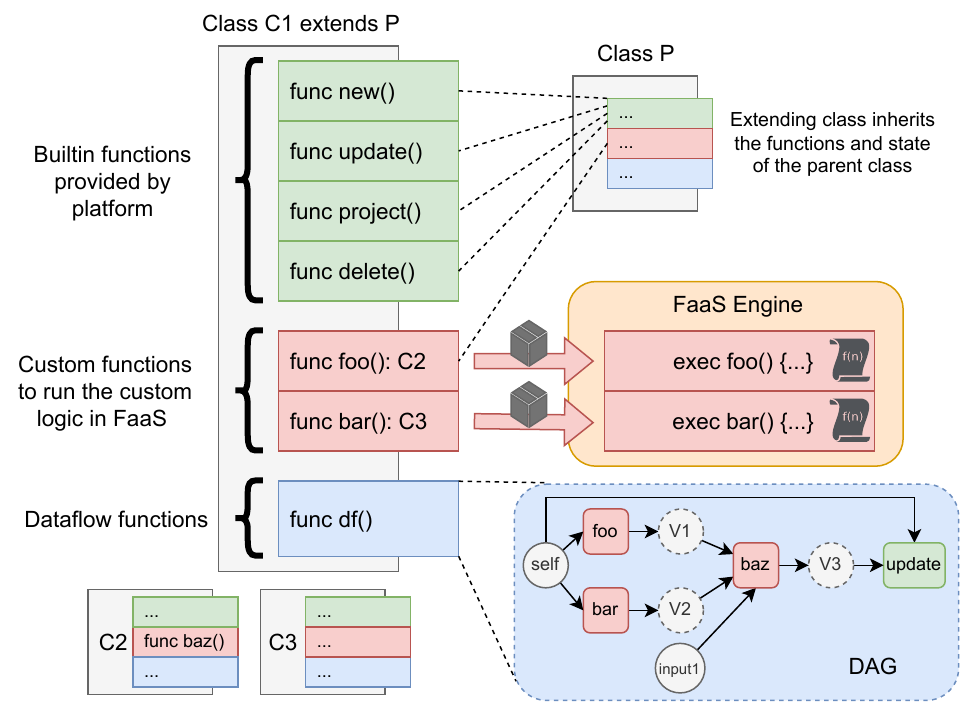}
  \caption{\small{Different types of functions supported by OaaS.}}
  \label{fig:class_representation}
  \vspace{-4mm}
\end{figure}

As shown in Figure~\ref{fig:class_representation}, OaaS supports three function types. First, \textit{built-in} functions that are provided by the platform. These functions could be the standard functions such as \texttt{CRUD} (create, read, update, and delete), which are the common data manipulation operations. The platform manages the execution of these functions without intervention from the developer. Second, \textit{custom} (a.k.a. \textit{task}) functions that are developed by developers (OaaS users) to provide their business logic. To handle the invocation of these functions, OaaS employs existing FaaS engines in its underlying layers to exploit their auto-scaling and scale-to-zero capabilities. Third, \textit{dataflow} (macro) functions are defined as a DAG representing the chain of invocations to objects.

As an example of package definition, Listing~\ref{lst:cls_exp} represents a declaration example for a package that includes one class called \texttt{video} that has a state named \texttt{mp4} (Line 6), \textit{built-in} function named \texttt{new} (Line 9), and \textit{custom} function named \texttt{transcode} (Line 1). The state \texttt{mp4} refers to video data that is unstructured data. The class has a public \textit{custom} function called \texttt{transcode}. The definitions of the \textit{custom} function are declared in Lines 15---17. The \texttt{type} of a function (Line 16) can be a \texttt{task} (or a \texttt{macro}, as noted earlier). This function creates another object instance of type \texttt{video} as an output. Line 17 declares the container image URI for executing function code.

\begin{minipage}{0.95\linewidth}
 \linespread{0.7}
\begin{lstlisting}[
    language=yaml, 
    label={lst:cls_exp}, 
    caption=\small{An example simplified script that declares \texttt{multimedia} package with a \texttt{video} class, and a \texttt{transcode} function for it in the \texttt{YAML} format. }
]
name: multimedia
classes:
  - name: video
    stateSpec:
      keySpecs:
        - name: mp4
          access: PUBLIC     
    functions:
      - function: new
        access: PUBLIC
      - function: transcode
        access: PUBLIC
        outputCls: .video
functions:
  - name: transcode
    type: TASK
    image: transcode-py:latest
    ...
\end{lstlisting}
\vspace{-1mm}
\end{minipage}



\section{Oparaca: A Platform for the OaaS Paradigm}
\label{sec:oprc}

\subsection{Design Goals}
\label{sec:oprc_goals}

The Oparaca platform is designed with its foundational goal of providing object abstraction with two additional design goals: \textit{backward compatibility} and \textit{extensibility}. \textit{\underline{first}}, while OaaS simplifies cloud-native application development, it is not always a replacement for FaaS; thus, Oparaca supports stateless FaaS and direct data access to storage systems. \textit{\underline{Second}}, for extensibility, Oparaca decouples the control plane from the execution plane, allowing the execution plane to operate independently via standardized APIs. This platform-agnostic design accommodates various execution planes optimized for specific use cases, such as latency-constrained function calls~\cite{singhvi2021atoll} or access to hardware accelerators~\cite{yang2022infless}.

\subsection{Overview of the Oparaca Architecture}
\label{sec:oprc_arch}

\begin{figure}
  \centering
  \includegraphics[width=0.95\linewidth]{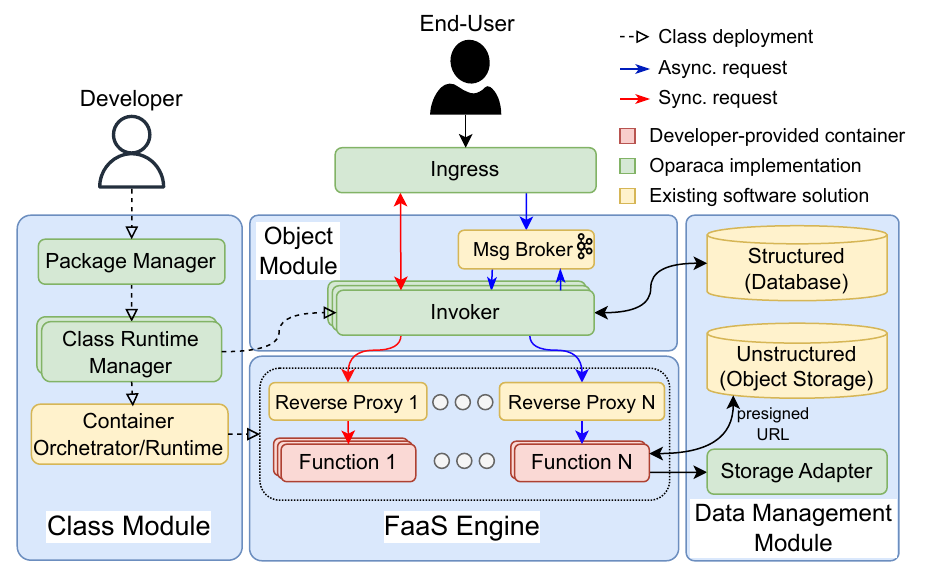}
  \caption{\small{A bird-eye view of the Oparaca architecture. Dashed lines show actions of the developer defining classes and objects, and solid lines show actions using objects and invoking functions. 
  }}
  \label{fig:oaas_architecture}
  \vspace{-4mm}
  
\end{figure}

The Oparaca platform is designed based on multiple self-contained microservices that communicate within a serverless system. Figure \ref{fig:oaas_architecture} provides a birds-eye view of the Oparaca architecture that is composed of five modules: 
\begin{itemize} [leftmargin=*]
    \item \textbf{Class Module} serves as the interface for developers to create and manage classes and their functions.
    \item \textbf{Object Module} serves as the cornerstone of Oparaca that has two main objectives: (a) providing the ``object access interface'' for the user application to access an object(s); and (b) offering the object abstraction while transparently handling function invocation and state manipulation.
    \item \textbf{FaaS Engine} is the underlying execution engine of Oparaca, which can be any existing FaaS system (e.g., Knative).
    \item \textbf{Data Management Module} is to manage object data persistence via employing database (e.g., document database) and object storage (e.g., S3-compatible storage). To bind these storages to the functions, the Invoker abstracts data access for structured data, while the Storage Adapter is employed to handle access to unstructured data in the object storage 
    \item \textbf{Ingress Module} whose purpose is to provide a single end-point for the user application.
\end{itemize}

Details of these modules, their interactions, and how they fulfill the consistency and fault-tolerance objectives (described in Section ~\ref{sec:intro}) are elaborated in the following subsections.
 
 \subsection{Class Module}
\label{sec:oprc_dep}
To define a new class and its functions in Oparaca, the developer defines them as a package definition and registers it to the \emph{Package Manager}, shown in Figure~\ref{fig:oaas_architecture}. Upon successful package validation by the Package Manager, the \textit{Class Runtime Manager} (termed \textit{CRM} for brevity) performs the class registration process that includes two operations: 

\noindent(\textbf{a}) Informing the Object Module about the new/updated class. Upon receiving a class registration, the Object Module creates a handler instance to be prepared for handling object invocation. We elaborate on this process in Section \ref{sec:oprc_inv}. 

\noindent(\textbf{b}) Registering the custom functions of the new class in the FaaS engine for future invocation. Recall that we aim to make Oparaca agnostic from the underlying FaaS engine. We design Oparaca to host a dedicated CRM for each FaaS engine. Accordingly, a new FaaS engine can be integrated into Oparaca by simply plugging its dedicated CRM into the system. When a function registration event occurs, the corresponding CRM processes this event by translating the function configuration into the specific format for that engine (e.g., Knative) and forwards it. Consequently, the underlying FaaS Engine creates the actual function runtime to be invoked by the Object Module.

\subsection{Object Module and FaaS Engine}
\label{sec:oprc_inv}


\begin{figure}[tbp]
  \centering
  \includegraphics[width=0.95\linewidth]{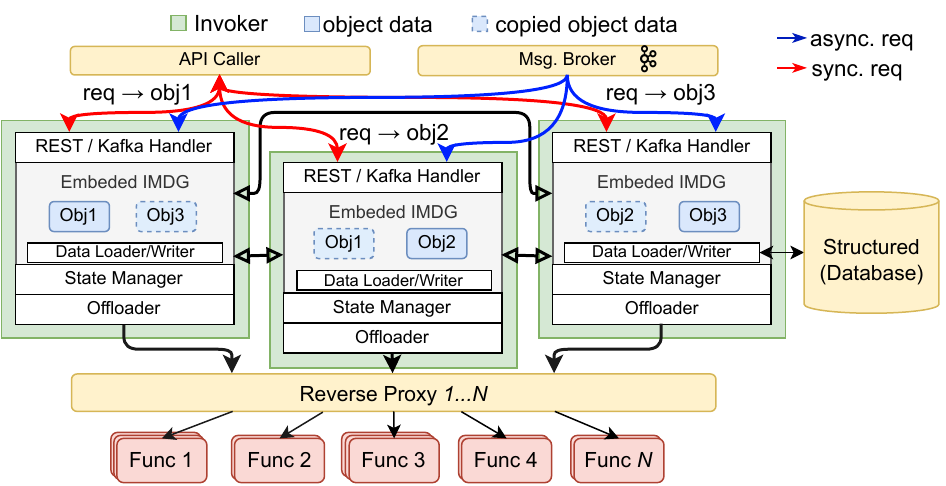}
  \caption{\small{The cluster of Invokers replicates and distributes object data across the cluster via IMDG with consistent hashing. The invoker offloads the invocation to a corresponding FaaS function.
  }}
  \label{fig:invoker}
  \vspace{-6mm}
\end{figure}

Recall that OaaS needs to support three types of functions: built-in, custom, and dataflow. Unlike built-in functions and dataflow functions that can be executed without the direct need of the FaaS engine, custom functions need to execute the developer-provided code on the FaaS engine. Thus, Oparaca requires a mechanism to utilize the FaaS engine to execute the custom function code while allowing it to access the object state transparently and with the minimum data transfer overhead. Needless to say, this mechanism also maintains the separation between the Object Module and the FaaS engine. 

To fulfill the above expectations, we design the object invocation mechanism in the Object Module by distinguishing between structured and unstructured states and managing it so that the data access overhead is minimized. We develop a hybrid approach that leverages the ``pure function'' technique for structured data access and the ``datastore abstraction'' technique for unstructured data access. The rationale of this design choice is that the unstructured state (i.e., BLOB) is usually large and expensive to transfer; hence, to maintain efficiency, the FaaS engine should retrieve the state directly from the object storage (e.g., S3) in a lazy, on-demand manner. This differs from the structured state, for which we include the state as an input argument to maintain a clear separation between the Object Module and the FaaS engine and let the FaaS engine maintain its statelessness.

In the Oparaca architecture (Figure~\ref{fig:oaas_architecture}), he mechanism for handling invocation and state management is managed by the \textit{Invoker} component. In particular, to offload the object invocation to the FaaS engine, Invoker bundles the request and the related structured object data as a ``\textit{task}'', as described in the next part, and passes it to the associated FaaS engine for execution.

\subsubsection{Task Generation in the Invoker}
\label{subsubsec:task_gen}
Upon receiving a function call, the Invoker bundles the invocation request and associated object data into the task and offloads it to be executed on the FaaS engine. To further reduce the data transfer overhead of providing the object abstraction in the task generation process, we design Invokers to maintain the object data (i.e., state and metadata) in a distributed hash table~\cite{hassanzadeh2021dht}, thereby reducing the cost of data transfer in a scalable manner. As shown in Figure \ref{fig:invoker}, we equip each Invoker instance with an embedded in-memory data grid (IMDG)~\cite{zhang2015memory}. IMDG partitions the entire data space into multiple segments and distributes them across Invoker instances. The Invoker with IMDG determines the segment for a given object by consistent hashing of the object ID and assigns the object data to the selected segment. Similarly, to retrieve the object data, IMDG determines the owner of the data and then fetches it from the owner of the segment in one hop.


\subsubsection{Unstructured Data Accessing}
\label{subsubsec:unstruc_access}
To minimize the overhead of accessing unstructured data, Oparaca allows function code to access the unstructured data on-demand and directly through a \textit{presigned URL} and \textit{redirection} mechanism. The presigned URL is the specific HTTP URL that includes the digital signature in query parameters to grant permission for anyone with this URL to access the specific data without the secret token. When a function needs to access the unstructured data, it sends an HTTP request to the storage adapter to receive the redirection response that points to the presigned URL of specific state data. Then, the function code can fetch the content directly from object storage via the given presigned URL. In addition to minimizing the overhead, using the presigned URL is important in protecting the function container from unauthorized access to other objects' data by analyzing their URL patterns.

\subsubsection{Task Completion}
\label{subsubsec:task_completion}
After the FaaS engine completes the task, it sends the task completion data to the Invoker to update the state. If the function reports a failed task, the state remains unchanged. Otherwise, the Invoker updates the object data in IMDG and then writes it to the persistent database immediately or asynchronously.  If an invocation involves both structured and unstructured states, we use pure and datastore techniques together, which can potentially lead to ``state inconsistency challenges''. We address this challenge in section~\ref{sec:oprc_consisitency}.


\subsubsection{Synchronous and Asynchronous Invocation}
As mentioned in Section~\ref{sec:oprc_goals} and shown in Figure~\ref{fig:invoker}, we designed Oparaca to offer synchronous and asynchronous function invocations. In synchronous mode, the function is executed immediately upon invocation and returns the result to the caller. Meanwhile, in asynchronous mode, the invocation ID is provided to the caller as a reference so they can check the invocation result later. The request is placed into the message broker to be reliably processed at a later time. To accommodate both modes, the Invoker utilizes the handler instance to accept the invocation request for either the REST API (synchronous) or the message broker (asynchronous). Subsequently, the handler instance forwards the request to be processed in the same way by the other part of the Invoker.

\begin{figure} [tbp]
  \centering
  \includegraphics[width=0.95\linewidth]{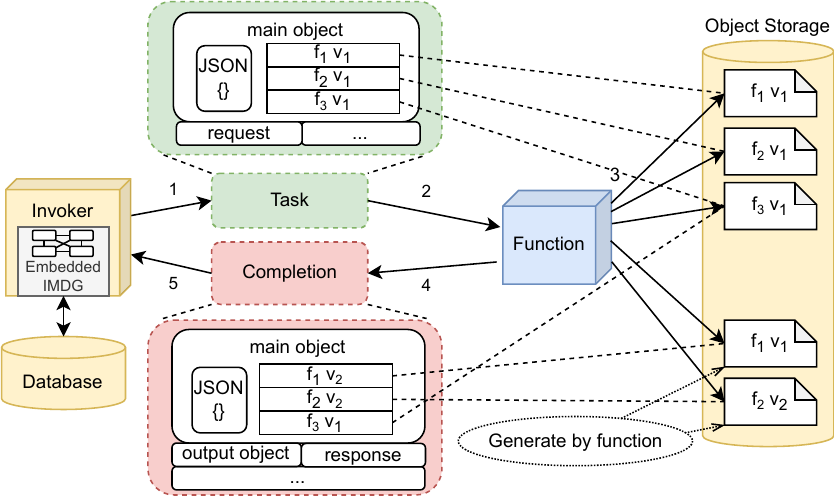}
  \caption{\small{The process of offloading invocation task into the function runtime. Invoker bundles the request input and object state into a task and offloads it to the function to be executed. With \emph{fail-safe state transition}, when the function needs to update the file in object storage, it creates a new file and updates the corresponding version ID via the returning completion message. }}
  \label{fig:invocation}
  \vspace{-6mm}
\end{figure}

\subsection{Ingress Module}
\label{sec:oprc_serve}

To provide the end user with a single access point, we position the Ingress Module in front of the cluster of Invokers. Additionally, to minimize data movement, the Ingress Module is designed to be aware of the object data distribution through consistent hashing of DHT. This allows the Ingress Module to correctly forward the object invocation request to the Invoker that owns the primary object data. As a result, the designated Invoker is able to access the data in its memory.

\subsection{Resilience Measures of Oparaca}
\label{sec:oprc_consisitency}
Oparaca is prone to the data inconsistency problem that stems from both \textit{failure} and \textit{race} conditions. In this section, we describe the internal mechanisms of Oparaca designed to make it resilient against these conditions.

\subsubsection{Resilience against failure}
Data inconsistency from failure can happen if the system stops while performing multiple update operations, causing some of the update operations to be incompletely executed. The pure function model, used for structured data in Oparaca, is inherently immune to this problem because a function returns the modified state to the platform only when its execution is complete. Nonetheless, the datastore abstraction used for the unstructured data in Oparaca is still prone to the data inconsistency problem between the structured database and object storage.

Maintaining data consistency across two data storages implies guaranteeing both storages are either successfully updated or fail for the same invocation. Otherwise (i.e., if only one of them succeeds), it leads to data inconsistency. To overcome this problem, we develop the \emph{fail-safe state transition} mechanism that disregards the data update in the object storage if Invoker fails to update the structured part of the object data in the structured database. For that purpose, the mechanism uses a two-phase versioning scheme to keep track of the unstructured data. As shown in Figure \ref{fig:invocation}, in the first phase, the mechanism creates a version ID for each file (unstructured data) and keeps them as structured data (metadata of object data) to track the current version of the file. In the second phase, which occurs upon function completion, Invoker changes all version IDs associated with the updated files (unstructured data) and then writes them to IMDG and the structured database. 

For example, consider object $o_1$ that has file $f_1$ with the version ID $v_1$. Upon function invocation, $f_1$ is updated and written to the object storage with version ID $v_2$. After the execution, the Invoker must change the version ID from $v_1$ to $v_2$ and commit the new structured object data. If any operation fails within this process, the next invocation still loads $o_1$ with version ID $v_1$, as if the previous invocation never happened. In the last step, when the invocation is complete, the Invoker purges the old and unused versions of data.

\subsubsection{Resilience against race condition}
Race conditions in Oparaca can occur when multiple invocations modify the same object data simultaneously, resulting in potential data inconsistency. One way to prevent this issue is by using database transactions; however, this method lacks abstraction as it allows direct function code access to the database and is tightly dependent on the type of database. 
An alternative approach to avoiding race conditions is the cluster-wide pessimistic locking mechanism to synchronize the locking state for all invokers. Nevertheless, this approach necessitates additional network communication to coordinate the locking state, which can lead to scalability issues. Alternatively, we develop an improved version of this mechanism, called ``localized locking,'' which relies on consistent hashing to direct the invocation request to the invoker that owns the primary copy of the targeted object data. Each invoker will only need to lock the object locally without additional network communication, making it more scalable than the cluster-wide version. Additionally, our localized locking approach guarantees that requests to the same object are executed in the arrival order, which is necessary in certain use cases where order matters, such as seat reservations. This is difficult to achieve with cluster-wide locking.

\subsubsection{Failure recovery in Oparaca}
\label{sec:oprc_fault}
To further establish resilience against failures, Oparaca is equipped with a mechanism to self-recover from the failure. Broadly speaking, a function invocation failure can be recovered by simply retrying the invocation. However, this approach can cause data incorrectness owing to the execution of the function more than once. The retrying approach could be undesirable for synchronous invocations because the failure can be handled on the client side. For asynchronous invocations, however, we need to guarantee that any invocation is only executed \textit{exactly once}.  

To achieve the \textit{exactly-once} guarantee, we have to prevent three sources of the problem that are: (a) losing messages, (b) duplicating messages, and (c) processing messages more than once. Message brokers with stable storage (e.g., Kafka~\cite{kreps2011kafka}) have features that can be leveraged to address these problems. To solve the first problem, upon failure occurrence, the Invoker can detect and reprocess the incomplete request using an offset number that is automatically generated by the message broker. The offset number is the auto-incremental number based on the message's arrival order and can be used to track the message's position in the queue. The second problem of producing duplicated request messages can be resolved using the message broker's ``idempotent producer'' feature. 

However, the message broker cannot completely address the third problem. That is, the Invoker can process the same invocation request more than once when the message broker has not acknowledged the completed one before the system failure occurs. We prevent this problem by tracking the offset number of the last processed request and adding it to each object metadata. In this manner, before processing an invocation request, Invoker checks the offset number of the target object to see if it is lower than the offset number of an incoming request. When the condition is met, the Invoker can detect that it has not been processed and perform the normal operation. Otherwise, it must be skipped to avoid reprocessing.

\subsection{Dataflow Abstraction in Oparaca}
\label{sec:oprc_dataflow}

To offer a high-level abstraction to declare a workflow, Oparaca provides the dataflow abstraction as a built-in feature that enables developers to declaratively define the invocation steps as a directed acyclic graph (DAG) in a domain-specific language (DSL) with YAML format. In every step, the developer can declare the output of each invocation as a temporary variable within the workflow. Then, the next invocation can use the temporary variables from previous steps as the input or target to call the function. Upon registering a dataflow function by the developer, Invoker constructs the DAG by having the invocation step as the edge and the objects as nodes.

Upon calling the dataflow function, one of the Invokers takes on the role of orchestrator, similar to the orchestrator pattern~\cite{richardson2018microservices} in microservices.  It breaks down the dataflow into multiple lower-level invocations and forwards them based on the topological order of DAG. Using consistent hashing, the invoker can determine the address of the target object and send the request directly to another Invoker that holds the target object. When each step is completed, the orchestrator keeps track of the intermediate dataflow state to transparently operate the data exchange between invocation steps. With the orchestrator pattern, the dataflow control logic is centralized into a single invoker, simplifying the management, monitoring, and error-handling implementation. 

When using the orchestrator pattern, the exact-once guarantee may be compromised because the object data is stored separately from the dataflow state. If the guarantee is needed, Oparaca allows flagging all invocation steps as immutable. Upon handling the dataflow request, Oparaca can generate the output ID in advance for each step, making each step of dataflow execution idempotent and safely re-executable.



\section{Discussion}\label{sec:discus}

\subsubsection{Security}
Certain security measures can be implemented in Oparaca to strengthen it against potential attacks. 
The \emph{first} measure is to reduce the attacking surface by limiting the necessary inbound traffic to the function container. As the function container is only accessed by the Invoker, the traffic policy can be configured to block inbound traffic except from the Invoker. The \emph{second} measure is to avoid reusing secret tokens. To prevent the function container from accessing out-of-context data via analyzing the URL path, we use the presigned URL mechanism for object storage. Thus, object storage in Oparaca is more secure than in FaaS, where the same secret key is used for every request. To secure the storage adapter, we can make the Invoker generate a unique secret token for each task, and every request for the storage adapter must be authenticated via the secret token. 

\subsubsection{Multi-tenancy}

The primary concern of multi-tenancy is ensuring data and resource isolation. The fundamental idea is to prevent sharing classes and functions among tenants. Since custom functions are offloaded and executed in a FaaS engine that provides strong isolation---with no shared functions---the execution environment is effectively contained within the FaaS engine. 
Regarding the Invoker and data management module, it is possible to share these components, as the data is stored separately in each class. However, depending on the billing model and isolation requirements, we can enhance security and resource isolation by separating these components for each tenant.

\subsubsection{Cold Start}
The developer functions and the Oparaca components can benefit from scale-to-zero to reduce the cost when there is no usage. However, this has the side effect of causing more cold starts. Since Oparaca components are shared across functions, we can effectively keep it warm to eliminate the additional cold start impact. In such a case, the cold start performance entirely depends on the underlying serverless execution engine.

\section{Performance Evaluation}
\label{sec:evltn}


\begin{figure*}[ht]
    \centering
    \subfloat[Video transcoding (sync)]{\includegraphics[width=0.33\textwidth]{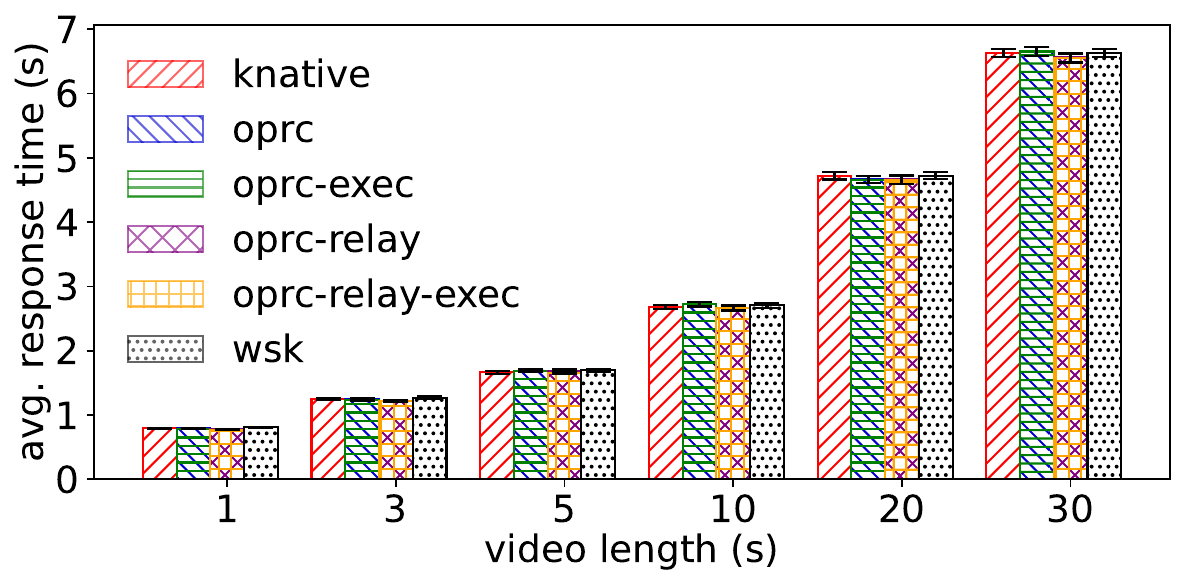}\label{fig:evlt:state_video_sync}}
    \hfill
    \subfloat[Text concatenation (sync)]{\includegraphics[width=0.33\textwidth]{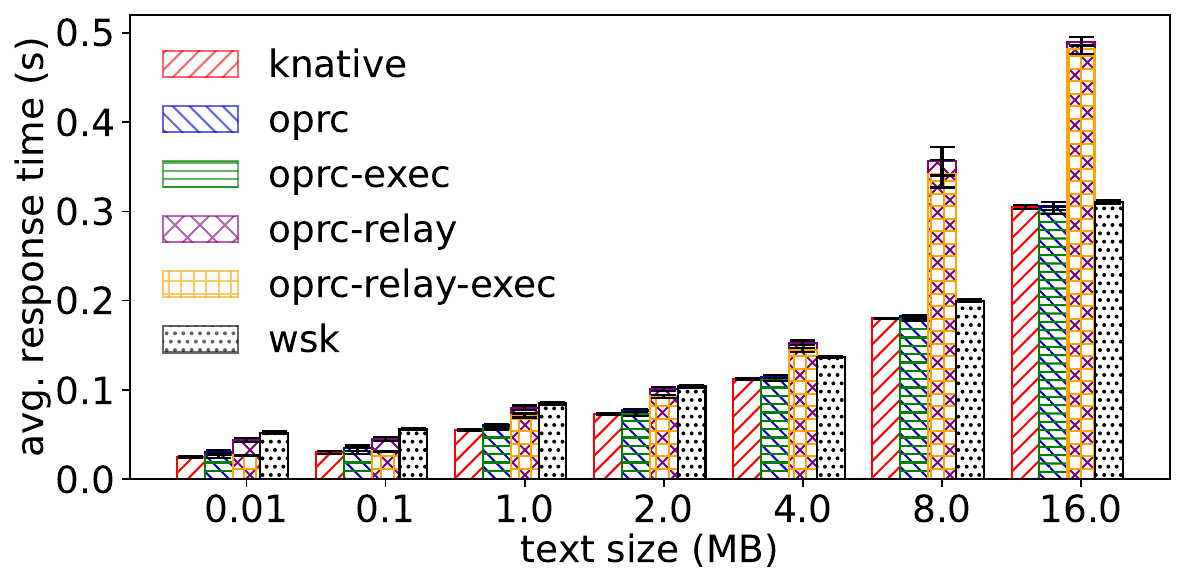}\label{fig:evlt:state_concat_sync}}
    \hfill
    \subfloat[JSON update (sync)]{\includegraphics[width=0.33\textwidth]{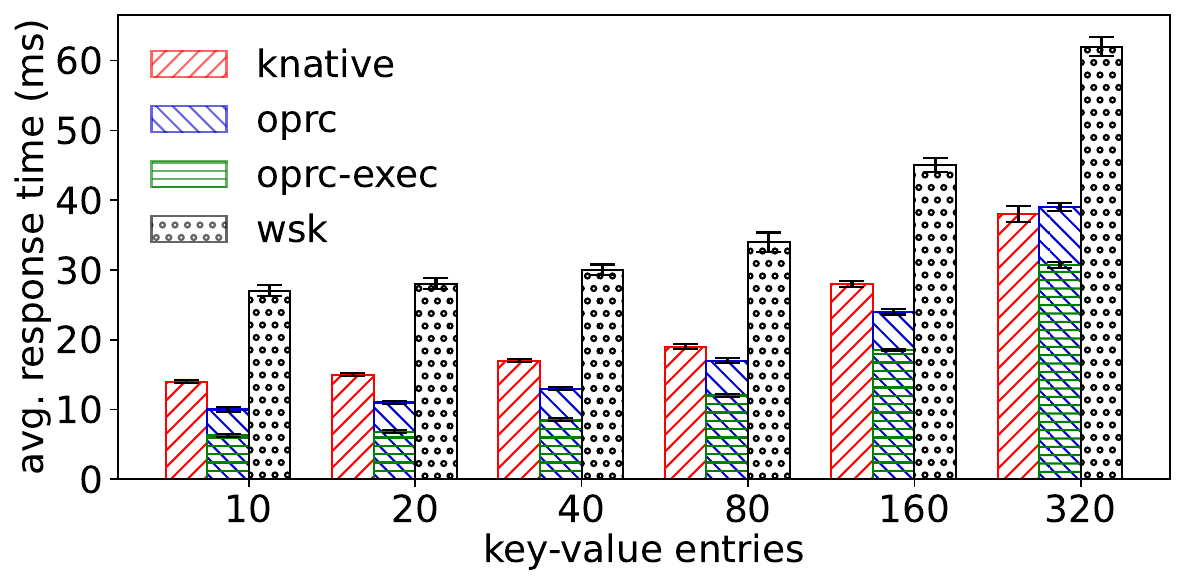}\label{fig:evlt:state_json_sync}}  

    \medskip

    \subfloat[Video transcoding (async)]{\includegraphics[width=0.33\textwidth]{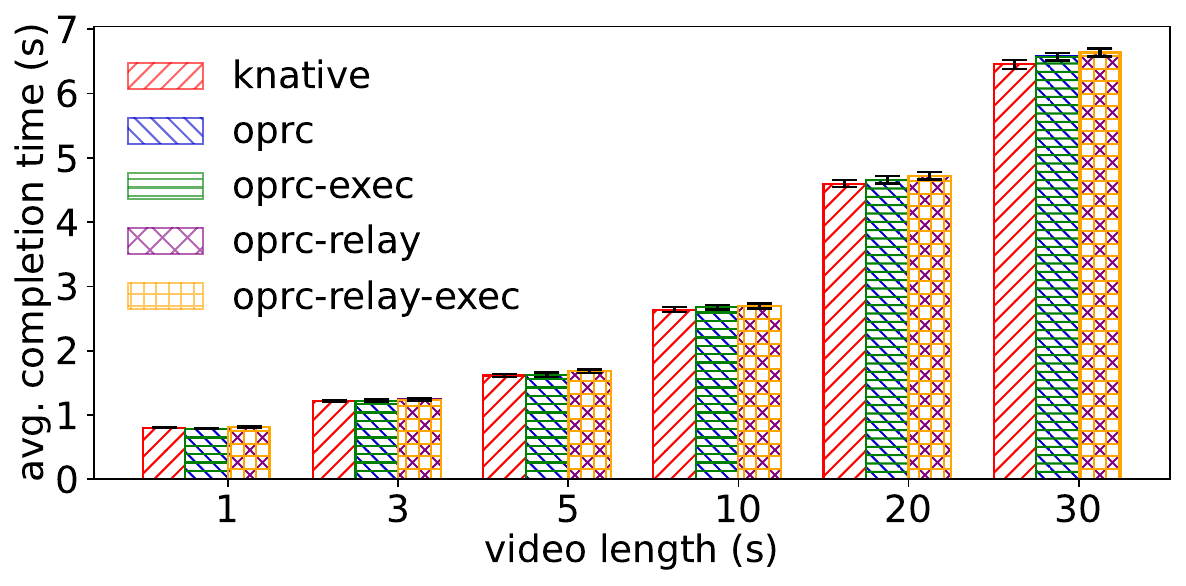}\label{fig:evlt:state_video_async}}
    \hfill
    \subfloat[Text concatenation (async)]{\includegraphics[width=0.33\textwidth]{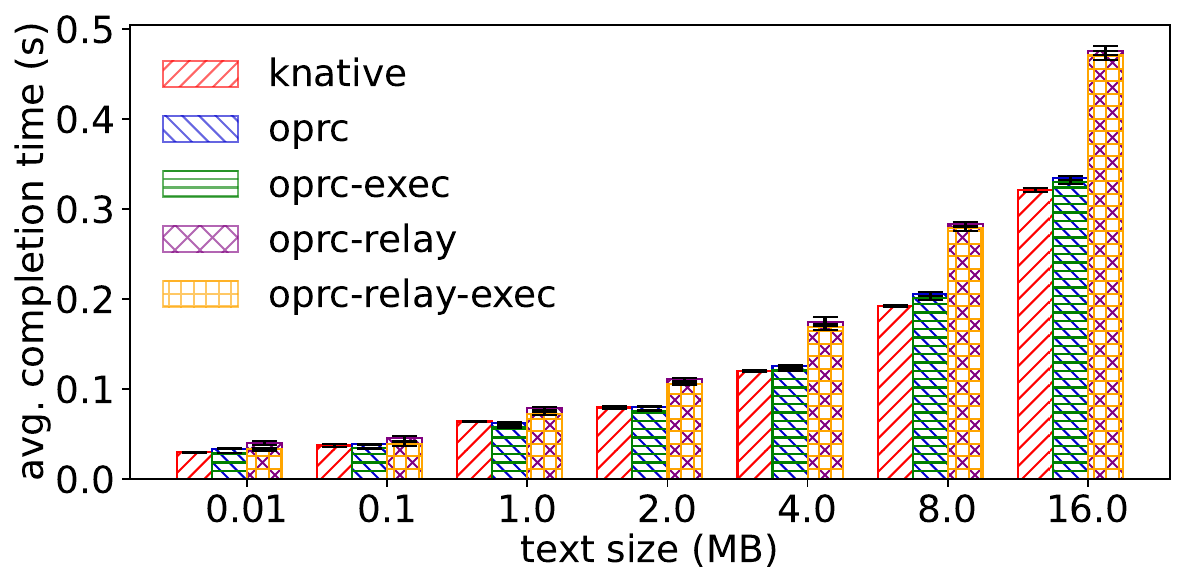}\label{fig:evlt:state_concat_async}}
    \hfill
    \subfloat[JSON update (async)]{\includegraphics[width=0.33\textwidth]{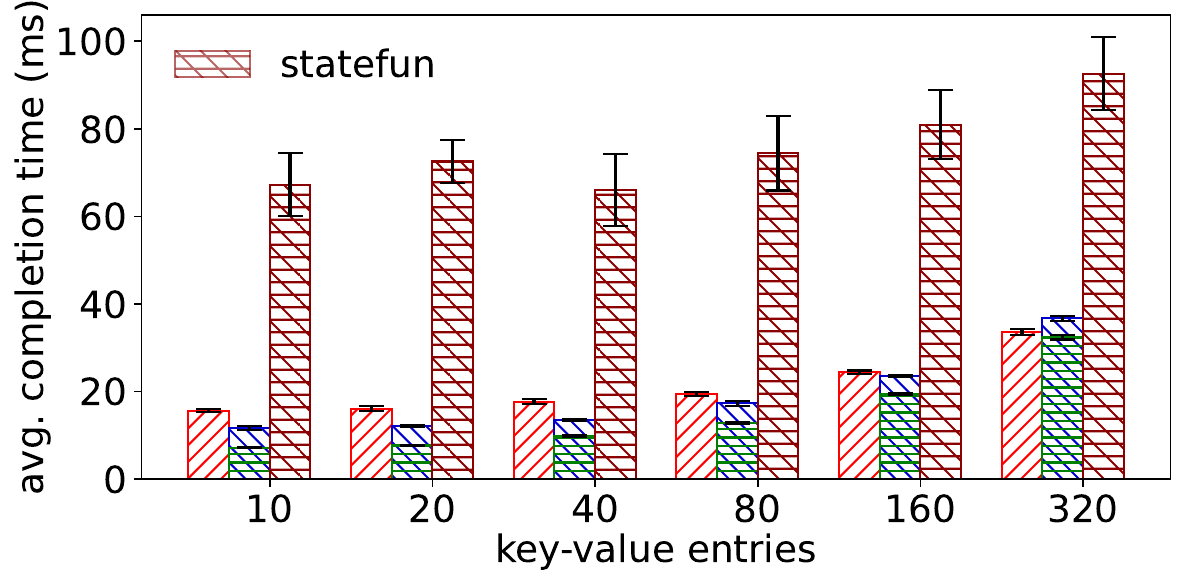}\label{fig:evlt:state_json_async}}  
  
    \caption{\small{
    The average execution time of functions for objects with various state sizes in synchronous and asynchronous invocations. Two versions of Oparaca are examined: the full version and the version without URL-redirection (\textit{oprc-relay}). We also capture the time used by the internal Knative in both Oparaca versions and show them with the suffix \emph{-exec} and plot them in the same bar as their Oparaca version.
    }}
    \label{fig:evlt:state}
\end{figure*}

\subsection{Experimental Setup}

We deploy the Oparaca platform on 4 machines of Chameleon Cloud \cite{chameleon_cloud}, each with 2 sockets of 24-Core Intel(R) Xeon(R) Gold 6240R CPU processors that collectively have 192 cores, 768 GB memory, and SSD SATA storage. We set up the Kubernetes cluster, which includes 15 VMs with 16 vCPUs and 32 GB of memory. We made another 2 VMs for the S3-compatible storage (Minio~\cite{minio}) for unstructured data and ArangoDB (\cite{arangodb}) for structured data. Oparaca is implemented using Java with Infinispan~\cite{infinispan} for IMDG.  

\noindent\textbf{Baselines.}
We configure Apache Flink Stateful Function (StateFun) \cite{statefun}, OpenWhisk \cite{openwhisk}, and Knative \cite{knative} to serve as the baselines. Unlike Oparaca and OpenWhisk, which focus on API calls and event handling, StateFun is an open-source stateful serverless system focusing on stream processing. Because StateFun does not manage the function worker instances out of the box, we configure Knative to complement it. OpenWhisk and Knative are popular open-source stateless FaaS platforms that we use to represent the state management done by the developer. 

We used Gatling\cite{gatling} for load generation and implemented three applications to serve as the workload. First is the video transcoding function, which utilizes FFmpeg~\cite{ffmpeg}, a CPU-intensive application. The second is a text concatenation function that concatenates the content of a text file (state) with an input string. This function represents a highly IO-intensive workload. Third is the JSON update function, which uses only structured data in JSON and is used to insert key-value pairs into the JSON state data randomly. The remaining workload characteristics are specific to each experiment and are explained in the respective sections. All three functions are implemented in the Python language. 

\subsection{Analyzing the Imposed Overhead of Oparaca}
\label{sec:evltn:ovh}
The abstractions provided by Oparaca are not free of charge and introduce some time overhead to the applications using these abstractions. In this experiment, our aim is to measure this overhead and see how the efficient design of Oparaca can mitigate this overhead. The latency of a function call is the metric that represents the overhead. We mainly study two sources of the overhead: (a) The \emph{state data size} that highlights the overhead of OaaS in dealing with the data, and (b) The \emph{concurrency of function calls} that highlights the overhead of the Oparaca system itself.


\begin{figure*}[ht]
    \centering
    \subfloat[Video transcoding function (sync)]{\includegraphics[width=0.33\textwidth]{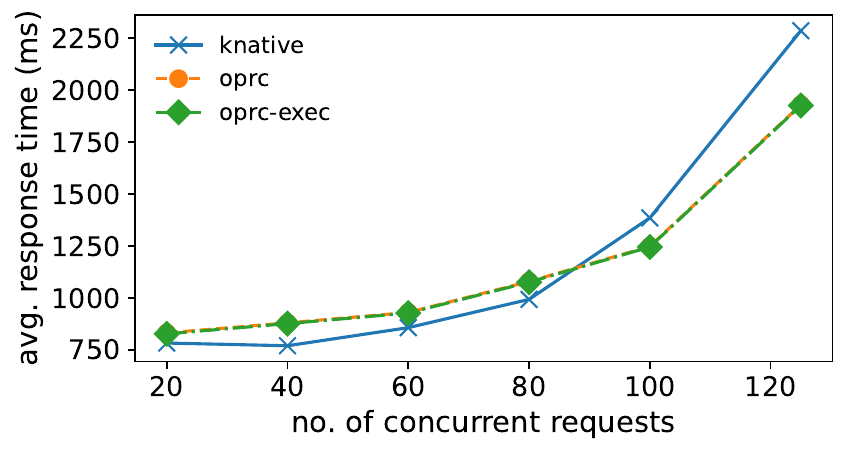}\label{fig:evlt:ccr_video_sync}}
    \hfill
    \subfloat[Text concatenation function (sync)]{\includegraphics[width=0.33\textwidth]{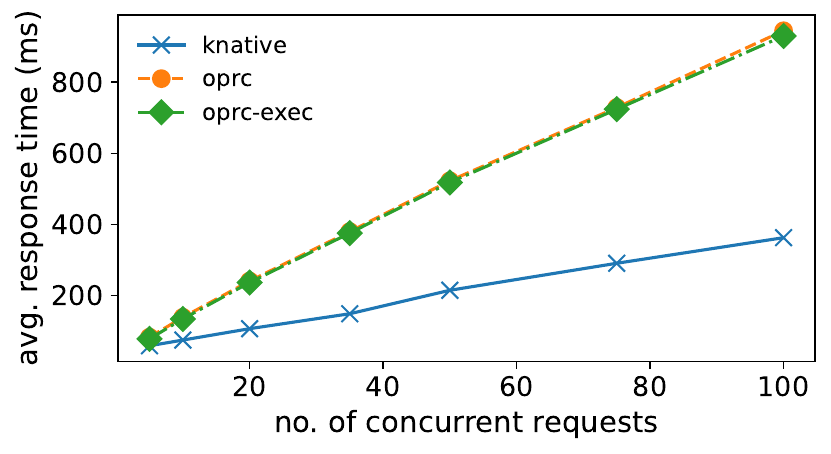}\label{fig:evlt:ccr_concat_sync}}
    \hfill
    \subfloat[JSON update function (sync)]{\includegraphics[width=0.33\textwidth]{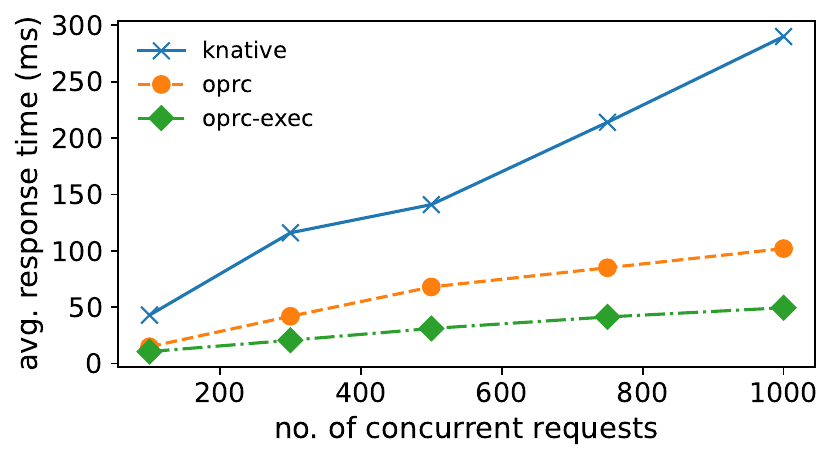}\label{fig:evlt:ccr_json_sync}}

    \medskip
    
    \subfloat[Video transcoding function (async)]{\includegraphics[width=0.33\textwidth]{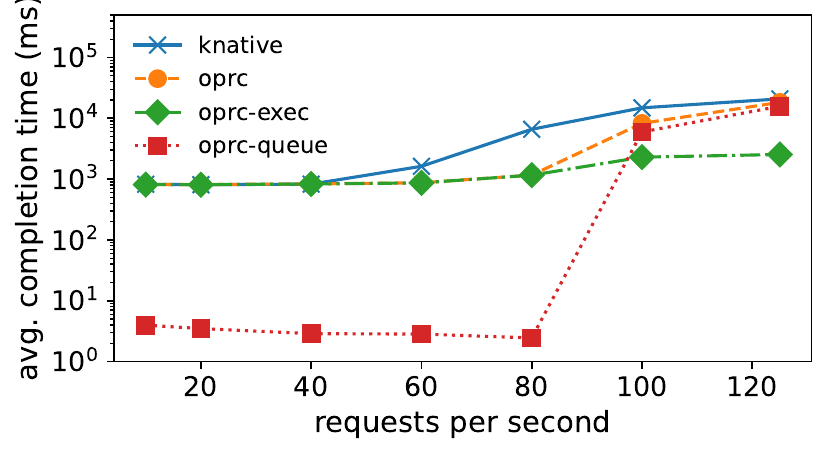}\label{fig:evlt:ccr_video_async}}
    \hfill
    \subfloat[Text concatenation function (async)]{\includegraphics[width=0.33\textwidth]{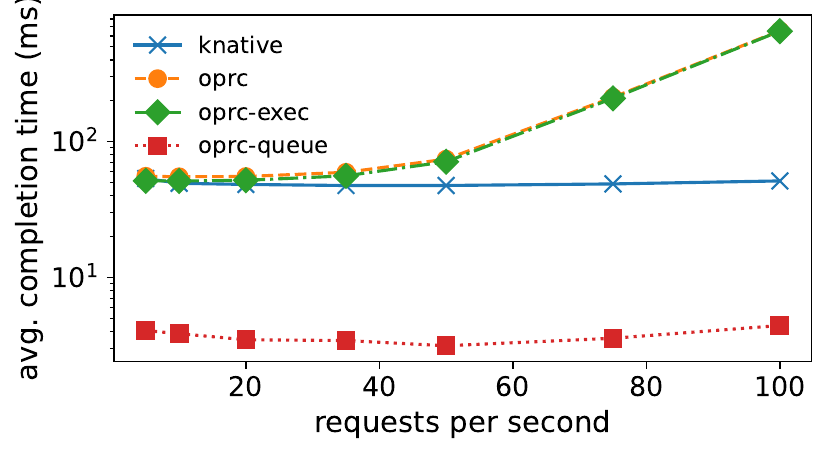}\label{fig:evlt:ccr_concat_async}}
    \hfill
    \subfloat[JSON update function (async)]{\includegraphics[width=0.33\textwidth]{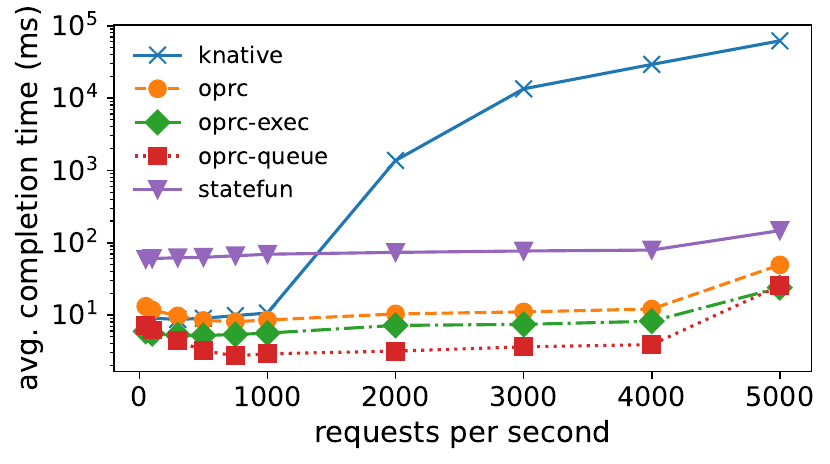}\label{fig:evlt:ccr_json_async}}
  
  \caption{\small{The average completion time of functions upon varying the rate of incoming requests in synchronous and asynchronous invocations. \textit{oprc-queue} is the queuing time that requests stay within the message queue}}
  \label{fig:evlt:ccr}
  \vspace{-5mm}
\end{figure*}

\noindent\textbf{The impact of changing the state size} is shown in Figure~\ref{fig:evlt:state}. To generate objects with various state sizes, we increased the input video length from 1---30 seconds. To remove the impact of video content on the result, the longer videos were generated by concatenating the same 1-second video. Similarly, the text files are from 0.01---16 MB. For the JSON object, the key and value sizes are 10 and 40 bytes, respectively, and the number of key-value pairs varies from 10---320 pairs. 
To concentrate only on the overhead of data access and avoid other sources of overheads, we configure Gatling to assign only one task at a time and set it to repeat this operation 100 times. To analyze the improvements offered by the URL redirection, we examine two versions of Oparaca: the full version (expressed as \emph{oprc}) and without URL redirection (expressed as \emph{oprc-relay}). The horizontal axes represent different state sizes for video, text, and JSON, respectively, and the vertical axes represent the average response/completion time (latency).

In Figure \ref{fig:evlt:state}, the average task execution time increases for larger state sizes. For the video transcoding function, all of the platforms perform with similar latency, which is expected because of the compute-intensive nature of the video transcoding that dominates the completion time. In a text concatenation function, however, Knative performs slightly better than Oparaca because of the overhead of unstructured state access by the redirection of the presigned URL. However, if we compare Oparaca with another version that uses a relay mechanism to provide the state abstraction, it performs much lower than its alternative with an average of 30\% lower response time. Lastly, we can see all the described trends happen similarly for synchronous and asynchronous request types. 

In the JSON update function (Figures~\ref{fig:evlt:state_json_sync} and \ref{fig:evlt:state_json_async}), Oparaca can perform with lower latency than Knative because the function does not need to fetch the object data from the database because of the pure function semantic. Nevertheless, Knative can catch Oparaca by increasing the key-value entries to 320. The reason is that the gain from eliminating the database connection is surpassed by the overhead of moving the data to the function code for larger records. OpenWhisk and Knative have the same pattern because both of them are FaaS, but OpenWhisk performs significantly worse. In Figure~\ref{fig:evlt:state_json_async}, the Statefun shares the same pattern with Oparaca with the consistent gap because it also relies on local storage to keep the function state without the need to fetch the data from the database. We also observe that Statefun performance degraded compared to our initial results \cite{lertpongrujikorn2023object}. This is because the storage hardware being used for the experiment has a lower through, which impacts the performance of Statefun.

\begin{figure*}[ht]
    \centering
    \begin{minipage}[htbp]{0.63\textwidth}
        \subfloat[Speedup results by horizontal scaling]{
        \includegraphics[width=0.51\textwidth]{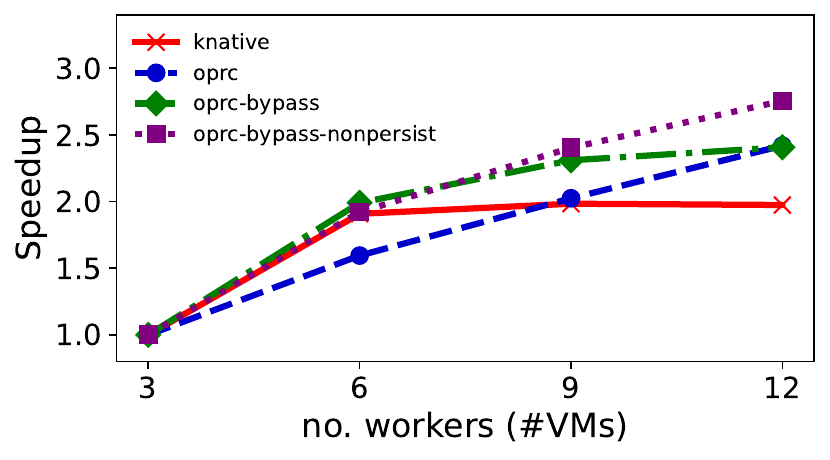}\label{fig:evlt:spd_json}
        }
        \hfill
        \subfloat[Throughput results from horizontal scaling]{
        \includegraphics[width=0.48\textwidth]{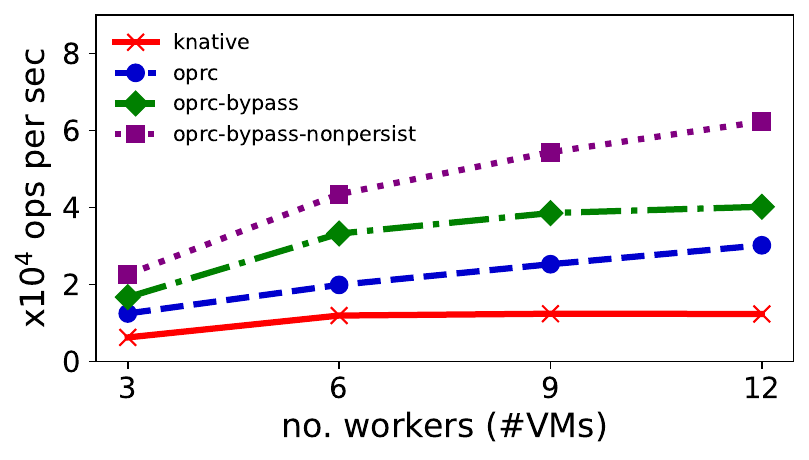}\label{fig:evlt:rps_json}
        }
        \caption{\small{Evaluating the scalability of the OaaS platform against other baselines. 
        }}
        \label{fig:evlt_scl_up}
    \end{minipage}
    \hspace{2mm}
    \begin{minipage}[htbp]{0.33\textwidth}
        \includegraphics[width=0.98\linewidth]{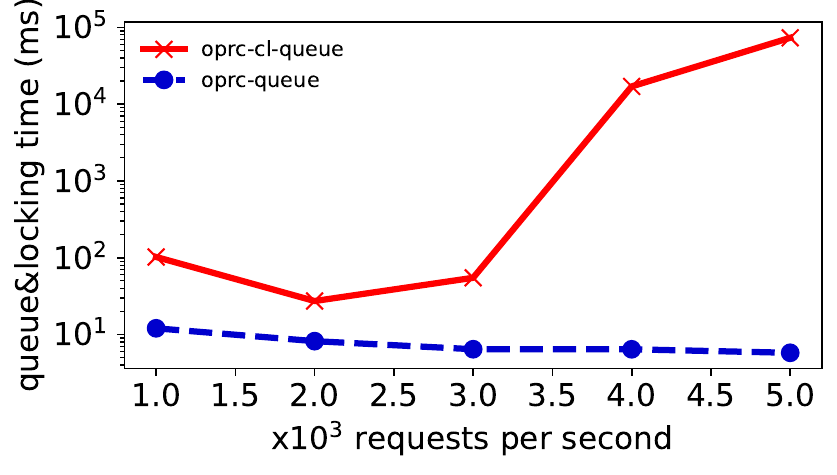}
        \caption{\small{Evaluating the performance of localized locking compared to cluster-wide locking}}
        \label{fig:evlt:lock}
    \end{minipage}
    \vspace{-5mm}
\end{figure*}

\noindent\textbf{The impact of concurrent function invocations} on the Oparaca overhead is shown in Figures \ref{fig:evlt:ccr}. In synchronous invocation, we increase the number of concurrent invocations of the same function (horizontal axes), whereas, for asynchronous invocation, concurrency depends on the system implementation which cannot be forced directly; thereby, we use the request arrival rate to increase the concurrency of invocations. To remove the impact of any randomness, We disabled the auto-scaling and limited the number of worker instances to 6. We also exclude OpenWhisk from this section because the Python runtime in OpenWhisk does not support container-level concurrency. 

For the transcoding function (Figures \ref{fig:evlt:ccr_video_sync} and \ref{fig:evlt:ccr_video_async}), at the low concurrency levels ($<$ 80 invocations), Oparaca has average response times higher than Knative, but for the higher concurrency levels, the response time of Knative grows faster than Oparaca due to computing resource limitations. Oparaca doesn't need to fetch video file metadata, giving it an edge at high concurrency. In the concatenation function (Figures \ref{fig:evlt:ccr_concat_sync} and \ref{fig:evlt:ccr_concat_async}), however, this phenomenon does not happen. The difference is that text concatenation is IO-intensive and desires high network bandwidth. The overhead of unstructured data access overwhelms the performance gain from eliminating structured data fetching.  

For the JSON update function (Figures \ref{fig:evlt:ccr_json_sync} and \ref{fig:evlt:ccr_json_async}), Oparaca can effectively reduce the latency by eliminating the need to fetch from the database. In Figure~\ref{fig:evlt:ccr_json_async}, because Statefun also shares this invocation scheme and, therefore, offers less completion time than Knative. However, since it relies on local storage to keep the state, while Oparaca uses the memory, Statefun's completion time is higher than Oparaca's. 


 In sum, Oparaca improves performance by eliminating database fetching but adds overhead by accessing unstructured data for secure state abstraction. Depending on the workload, this can either improve or impair object function invocation performance. The overhead may outweigh I/O-intensive workloads, but Oparaca can improve latency by up to 2.27x compared to Knative for workloads without unstructured data.

\noindent
\colorbox{blue!10}{
\parbox{0.47\textwidth}{
\underline{\textbf{Takeaway}:} \emph{Object abstraction can be provided with an insignificant latency overhead for objects with only a structured state. The main object overhead occurs as a result of securing unstructured data access.}
}}

\vspace{-4mm}
\subsection{Scalability of the Oparaca Platform}

\label{sec:evltn:scl}


To study the scalability, we scale out the Kubernetes workers from 3---12 VMs, each with 16 vCPU cores (in total 48---192 vCPUs). We measured throughput and speedup metrics, focusing on the JSON update function, which does not rely on slow object storage, which becomes the bottleneck of this experiment. We measure the throughput by continually increasing the concurrency until the throughput stops growing (Figure \ref{fig:evlt:rps_json}). We assume three VMs as the base speedup=1, and the speedup of other cluster size is calculated with respect to the base value. Moreover, we add two other versions of Oparaca: first is \textit{oprc-bypass} that uses a standard Kubernetes deployment as its underlying function execution instead of Knative; Second is \textit{oprc-bypass-nonpersist} that does not persist the object data to the database to measure if Oparaca is not bottleneck by the database write operation. 

According to Figure~\ref{fig:evlt:spd_json}, the speedup of Knative plateaus after reaching 6 VMs. We realized that this plateau is attributed to the database write operation throughput bottleneck. Conversely, Oparaca exhibits the potential for higher speedup enhancement due to its reliance on the distributed in-memory hash table to consolidate data for batch write operations. This approach can boost maximum throughput by up to 3.27x when comparing \textit{oprc-bypass} with \textit{knative}. 

Figure \ref{fig:evlt:rps_json} shows that \textit{oprc-bypass} yields a higher throughput over the baseline Oparaca. This is because Oparaca sends task data through the Knative internal proxy to offload the task to Knative. While this setup allows for scale-to-zero functionality, bypassing these components leads to even higher throughput. Furthermore, by disabling the database writing operation, which is the bottleneck, \textit{oprc-bypass-nonpersist} can achieve even higher throughput. Although there isn't linear scalability due to the limitations of the database write performance, Oparaca significantly improves maximum throughput compared to traditional FaaS systems.

\noindent
\colorbox{blue!10}{
\parbox{0.47\textwidth}{
\underline{\textbf{Takeaway}}: \emph{In addition to offering a higher-level abstraction, Oparaca can improve the throughput and response time of its underlying Knative engine via reducing database operations, thereby, mitigating its bottleneck.
}
}
}
\vspace{-3mm}


\subsection{Performance of Localized Locking}

To analyze effectiveness of localized locking, we created a variation of Oparaca, called \textit{oprc-cl}, that has cluster-wide locking and evaluated it using a cluster of 12 Invokers while increasing the request arrival rate to measure the locking overhead. To generate requests involving the locking mechanism, we created multiple requests targeting the same object. 
From Figure~\ref{fig:evlt:lock}, the overhead of localized locking remains mostly constant, while the overhead of cluster-wide locking rises for higher request rates. 
The cluster-wide version does not exhibit this behavior, as the network communication overhead limits the throughput and hinders high invocation rates.

\vspace{-4mm}
\subsection{Case Study: Development Efficiency Using OaaS}
\label{sec:evltn:dev}

\begin{figure}[tbp]
  \centering
  \includegraphics[width=0.45\textwidth]{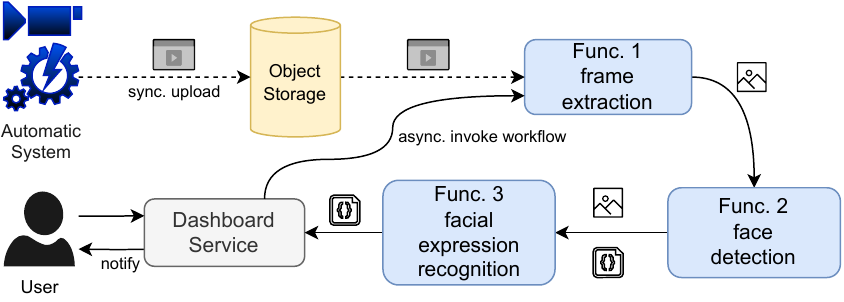}
  \caption{\small{The use case of developing a face expression recognition workflow for an input video.}}
  \label{fig:evltn:scn}
  \vspace{-5mm}
\end{figure}

In this part, we provide a real-world use case of object development using OaaS and its FaaS counterpart and then demonstrate how OaaS makes the development process of cloud-native serverless applications easier and faster. 

\vspace{1mm}
\noindent\textbf{\underline{Case Study \# 1. expression detection system.}} This case study is a video processing workflow that performs face detection and facial expression recognition. Figure \ref{fig:evltn:scn} shows the automatic system uploads the video file to the object storage to be processed by the workflow of functions. 
The workflow includes: \texttt{Func\_1} to split the video into multiple image segments; \texttt{Func\_2} to detect the face on each sample image frame; and \texttt{Func\_3} to perform facial expression recognition on the detected face image and generate a \texttt{JSON} format label. These functions must persist their output object so that the next function in the workflow can consume it.

\noindent\textbf{FaaS implementation.} The developer must implement the following steps: (a) Configuring cloud-based object storage and managing access tokens. (b) Implementing business logic to respond to trigger events. (c) Manage data within the functions that involve three steps: allocating the storage addresses, authenticating access to the object storage, and performing fetch and upload operations to the allocated addresses. 
Upon implementing these functions, the developer has to connect them as a workflow via a function orchestrator service. Finally, the dashboard service invokes the workflow upon receiving a user request and collects the results. 
	
\noindent\textbf{OaaS implementation.} The developer defines three classes, namely \texttt{Video}, \texttt{Image}, and \texttt{Expression}, in the form of the three following classes: 
(a) \texttt{Video} class with \texttt{frame\_extract()} functions; and a macro function, \texttt{df\_exp\_recog(detect\_interval)}, that includes the whole dataflow of function calls, with the requested sampling period as its input, and an \texttt{expression\_data} object as the output.
(b) \texttt{Image} class with the \texttt{face\_detect()} and \texttt{exp\_recognize()} functions. 
(c) \texttt{Expression} class that does not require any function. 
The dashboard service calls the \texttt{wf\_exp\_recognize(detect\_interval)} dataflow function directly using the object access interface and receives the \texttt{Expression} object as the output.
We note that the developer does not need to be involved in the data locating and authentication steps when developing the class functions because of the abstraction that OaaS provides.

\begin{figure}[t]
    \centering
    \subfloat[FaaS-based]{
        \includegraphics[width=0.9\columnwidth]{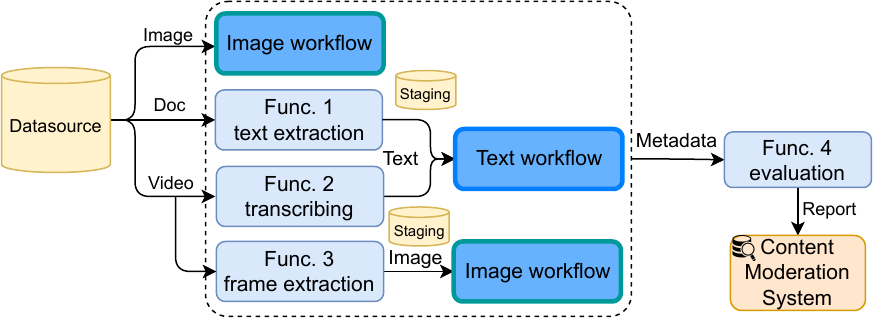}
        \label{fig:evlt:case-study2-faas}
    }
    \vspace{3mm}
    \subfloat[OaaS-based]{
    \includegraphics[width=0.9\columnwidth]{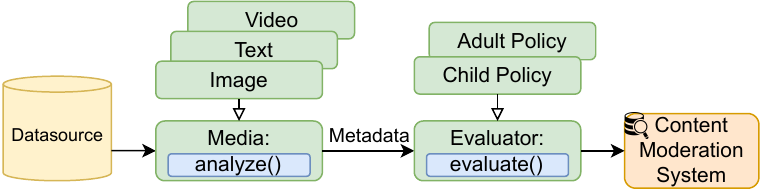}
        \label{fig:evlt:case-study2-oaas}
    }
    \caption{The automatic content moderation system.}
    \label{fig:evlt:case-study2}
  \vspace{-6mm}
\end{figure}

\vspace{1mm}
\noindent\textbf{\underline{Case Study \# 2. content moderation system.}} Moderating the content at scale in various formats, including image, document, and video. We first present how the application is deployed in FaaS, the limitations of this approach, and how OaaS can resolve those limitations.

\vspace{0.5mm}
\noindent\textbf{FaaS implementation.} To simplify complex multimedia processing workflows~\cite{aws-cms}, we abstract the workflow to extract the metadata from the image files as \texttt{Image workflow}, and the workflow to extract metadata from the text files as \texttt{Text workflow} as shown in Figure~\ref{fig:evlt:case-study2-faas}. Before using both workflows, the content must be pre-processed to extract raw data via using the pertinent FaaS functions: (a) \texttt{text extraction} to extract text from the document. (b) \texttt{transcribing} to extract text from the video. (c) \texttt{frame extraction} to sample image frame from the video. After feeding the data into both workflows to extract the metadata, we use the \texttt{evaluation} function to generate a report to the content moderation system.

The FaaS implementation has three main drawbacks: \textbf{(A)} developers must explicitly manage application state and data using separate storage services that increase the complexity. \textbf{(B)} even though the common workflow can be reused, the intermediate data management is not abstracted. That is, if the developer needs to separate/change the staging storage, it must be done manually. \textbf{(C)} functionalities may require numerous and heterogeneous FaaS deployments---e.g., requiring a separate workflow for each content type, where the Video format requires a more complicated workflow than the other types. These drawbacks complicate development, deployment, and management as the application evolves to handle various document types and integrates more functionalities and options (e.g., using multiple evaluation services instead of one).

\noindent\textbf{OaaS implementation.} To demonstrate the efficacy of OaaS, we transform the given FaaS-based solution into OaaS with minimal effort to resolve the aforementioned drawbacks.

\begin{itemize}[leftmargin=*, noitemsep, topsep=0.5pt]
\item\textbf{Workflow Construct.}
We encapsulate related FaaS functions and states into classes representing two key functionalities: \texttt{Media} to extract the metadata and \texttt{Evaluator} to evaluate metadata and report to the content moderation system. The two classes form the critical path of the application processing pipeline, as shown in Figure \ref{fig:evlt:case-study2-oaas}.

\item\textbf{Object Encapsulation.} We use inheritance and polymorphism to enhance software reuse by encapsulating FaaS functions and states in classes derived from two base classes. This approach hides the need for storage services behind object abstraction and allows their implementation to be managed in the cloud. It simplifies development, as developers only build the processing pipeline once in the base classes. They can then focus on specific functionalities in the derived classes, avoiding \textbf{redundant} implementation when adding new content types or evaluator services.
\end{itemize}

\noindent
\colorbox{blue!10}{
\parbox{0.47\textwidth}{
\textbf{\underline{Takeaway}:} \emph{The OaaS paradigm aggregates the state storage and the function workflow in the object abstraction and enables cloud-native dataflow programming. Thus, developers are relieved from the burden of state management, learning the internal mechanics of the functions and pipelining them. 
}
}
}
\vspace{-3mm}
\section{Conclusions}\label{sec:conclsn}

In this research, we presented the OaaS paradigm, which aims to simplify state and workflow management for cloud-native applications. Our prototype, Oparaca, supports both structured and unstructured data types, ensuring consistency through fail-safe state transitions. It also provides secure and low-overhead management for unstructured data using presigned URLs and redirection mechanisms. 
For structured data, it employs the pure function scheme to transparently manage application data to the developer code and using DHT and consistent hashing to scalably cache the object data and improve the data locality. To make the Oparaca fault-tolerant, we developed the \textit{exactly-once} and \textit{localized locking} schemes. To support cloud-native workflow, Oparaca enables declarative dataflow abstraction that hides the concurrency and synchronization concerns from the developer's perspective. The evaluation results demonstrate that Oparaca streamlines cloud-native programming and is ideal for use cases that require persisting the state or defining a workflow. Oparaca offers scalability with negligible overhead, particularly for compute-intensive tasks.
In the future, we plan to develop Oparaca to support application deployment across multiple data centers, streamlining large-scale application development.

\vspace{-12pt}
\section*{ACKNOWLEDGEMENT}
We would like to thank anonymous reviewers for their constructive feedback; and Chameleon Cloud for providing resources.
This project is supported by National Science Foundation (NSF) through CNS CAREER Award\# 2419588.

%
\bibliographystyle{plain} 
\balance
\bibliography{references}

\vspace{-10mm}
\begin{IEEEbiography}
[{\includegraphics[width=1in,height=1.25in,clip,keepaspectratio]{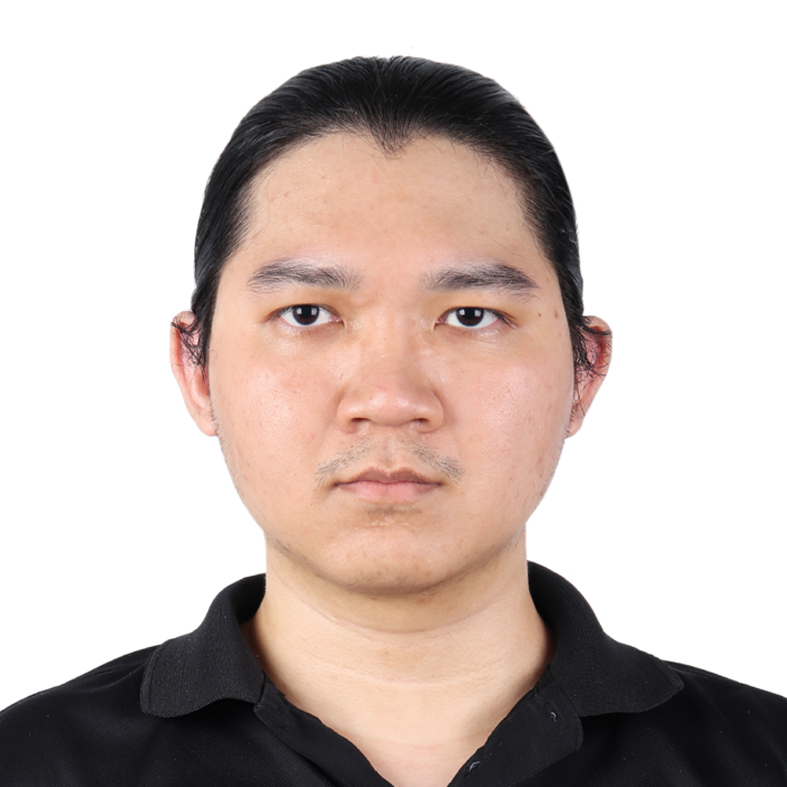}}]{Pawissanutt Lertpongrujikorn} received B.Eng. in computer engineering from Kasetsart University, Thailand in 2019. Currently, He is working toward his Ph.D. at HPCC Lab, University of North Texas. His research focuses on developing a cloud-native programming paradigm and serverless computing.
\end{IEEEbiography}

\vspace{-8mm}
\begin{IEEEbiography}
[{\includegraphics[width=1in,height=1.25in,clip,keepaspectratio]{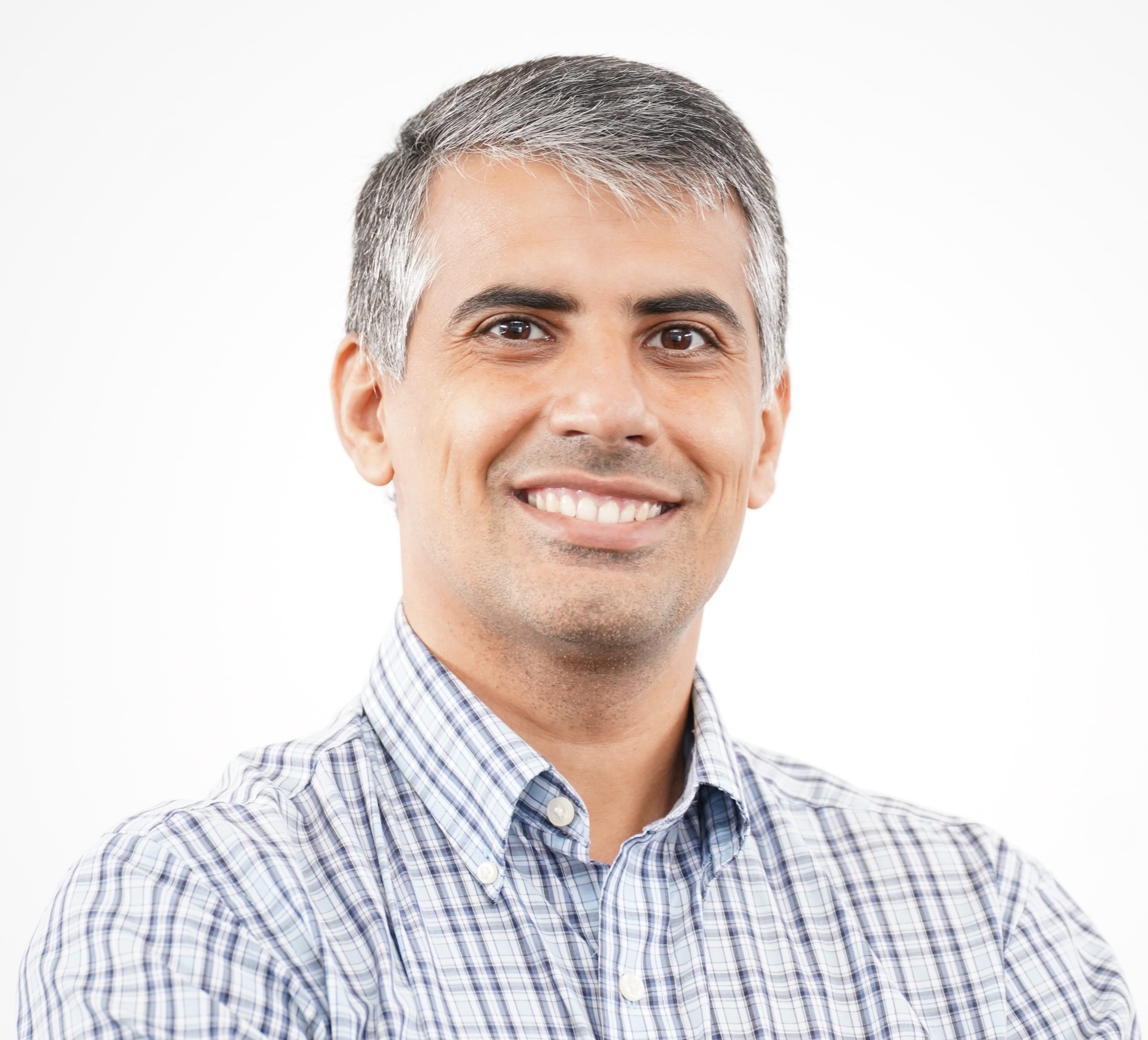}}]{Dr. Mohsen Amini Salehi} is an Associate Professor and the director of the \href{https://hpcclab.org}{HPCC} Lab, at the Computer Science and Engineering dpt., University of North Texas. His team focuses on democratizing cloud-native application development and building smart and trustworthy systems across edge-cloud. He is an NSF CAREER Awardee and, so far, he has had 11 research projects local and federal agencies. 
\end{IEEEbiography}

\end{document}